\documentclass{article}
\usepackage{amsmath}
\usepackage{amssymb}
\usepackage{amsmath}
\usepackage{amsfonts}
\usepackage{hyperref}
\usepackage[parfill]{parskip}
\usepackage{graphicx}
\let\UnmodifSec=\section
\renewcommand{\section}{\setcounter{equation}{0}\UnmodifSec}

\def\bR{{\bf R}}

\def\Im{\mathop{\rm Im}\nolimits}

\def\CC{{\cal C}}

\def\EE{{\cal E}}

\def\TT{{\cal T}}

\def\wh{\widehat}

\def\interior#1{\setbox1=\hbox{$#1$}\rlap{$#1$}\kern0.4\wd1\raise1.1\ht1%
\hbox{$\scriptstyle \circ$}}

\def\boxit#1#2{\setbox1=\hbox{\kern#1{#2}\kern#1}%
\dimen1=\ht1 \advance \dimen1 by #1 \dimen2=\dp1 \advance \dimen2 by
#1
\setbox1=\hbox{\vrule height\dimen1 depth\dimen2\box1\vrule}%
\setbox1=\vbox{\hrule\box1\hrule}%
\advance \dimen1 by .4pt \ht1=\dimen1 \advance \dimen2 by .4pt
\dp1=\dimen2 \box1\relax}
\def\endprf{\raise .5ex\hbox{\boxit{2pt}{\ }}}

\def\ifundefined#1{\expandafter\ifx\csname#1\endcsname\relax}

\def\beq{\begin{equation}}
\def\endq{\end{equation}}
\def\beqa{\begin{eqnarray}}
\def\endqa{\end{eqnarray}}

\begin{document}
\title{Scalar tachyons in the de Sitter universe}

\author{Jacques Bros$^1$, Henri Epstein$^2$ and Ugo Moschella$^3$ \\
$^1$Service de Physique th\'eorique - CEA. Saclay.
91191 Gif-sur Yvette.\\$^2$Institut des Hautes \'Etudes
Scientifiques, 91440 Bures-sur-Yvette.\\$^3$Universit\`a
dell'Insubria, Como and INFN Milano}
\maketitle

\begin{abstract}
We provide a construction of a class of local and de Sitter covariant 
tachyonic quantum fields which exist for discrete negative values of the 
squared mass parameter and which have no Minkowskian counterpart.
These quantum fields satisfy an anomalous non-homogeneous 
Klein-Gordon equation. The anomaly is a covariant field 
which can be used to select the physical subspace (of finite codimension) 
where the homogeneous tachyonic field equation holds in the usual form.
We show that the model is local and de Sitter invariant on the physical space.
Our construction also sheds new light on the massless 
minimally coupled field, which is a special instance of it.
\end{abstract}

The word  tachyon denotes a would-be  particle traveling faster than light.
Despite the fact that tachyons are believed to be unphysical,
they play an important role in many circumstances as for instance in 
bosonic string theory,
and the question about their existence and physical meaning
is far from being settled.
The commonly accepted interpretation is that the appearance of
tachyonic degrees of freedom in a theoretical model points towards an 
instability
of the model, and this instability should be
described by tachyonic fields as opposed to real faster-than-light particles; 
the instability
can be treated by invoking the so-called tachyon condensation.

Starting with the observation of distant supernovae in 1998, the
cosmological constant has taken a central place in the scene of
contemporary physics; since then many widespread beliefs have been
shattered and abandoned. A nonzero cosmological constant  may also be
responsible for a change in the understanding we have of tachyons,
as we do show in this letter. A nonzero cosmological constant
renders indeed possible the existence of a class of tachyonic fields
on the de Sitter universe that do not share the problems of their
corresponding Minkowskian counterparts, and on the contrary give rise
to acceptable free field theories.

We will discuss the $d$-dimensional de Sitter manifold
with unit radius, described as the hyperboloid
$X_d = \{x\in {\bf R}^{d+1}, \, x^2 = -1\}. $ On this manifold we
consider Klein-Gordon fields for general complex values
of the squared mass $m^2 = - \lambda(\lambda+d-1)$; tachyons
correspond to real negative squared masses.
Since we are dealing with free fields, the knowledge of a
two-point function solving the de Sitter Klein-Gordon equation
\begin{equation}
\Box w_\lambda(x_1,x_2) =  \lambda(\lambda+d-1) w_\lambda(x_1,x_2)\label{kg2p}
\end{equation}
w.r.t. both $x_1$ and $x_2$ allows for a complete reconstruction
of the field.
One would like to impose the following properties
on the two-point function:

\noindent$\bullet$ invariance under the global symmetry group:
for $X_d$, the latter is the de Sitter group $SO_0(1,d)$;

\noindent$\bullet$ locality, i.e. the vanishing of the commutator
\begin{equation}
c_\lambda(x_1,x_2) =  w_\lambda(x_1,x_2)- w_\lambda(x_2,x_1)\label{kgcomm}
\end{equation}
at spacelike separation  and

\noindent$\bullet$  positive-definiteness, to warrant a bona-fide
quantum mechanical interpretation.

The above requirements have proven to be incompatible for tachyonic
fields on the Minkowski manifold. Feinberg \cite{feinberg} considers 
tachyons as scalar fermions abandoning the locality condition 
while Schroer \cite{schroer}
constructs local and covariant tachyon fields which do not satisfy
the positive-definiteness condition and have no direct quantum
mechanical interpretation.

Even when the squared mass is positive, the above properties do not
uniquely prescribe the two-point function and one needs physical or
mathematical criteria to discriminate among the various
possibilities. The "preferred vacua" of free de Sitter QFT's, also
known as the Bunch-Davies vacua or "Euclidean vacua"
\cite{Th,ChT,GiH,N,BuD,Allen}, can be uniquely identified by the
property of maximal analyticity of their two-point functions
\cite{bmg,bros}, i.e. the latter arise as boundary values of
functions which are analytic in the cut domain \beq \Delta = \{
(z_1,z_2) \in X_d^{(c)} \times X_d^{(c)}  \, :\;(z_1-z_2)^2 = - 2 -2
\,z_1\cdot z_2 \not = c
> 0\}.
\endq
$X^{(c)}_d$ being the complex de Sitter hyperboloid $X^{(c)}_d =
\{z\in {\bf C}^{d+1}, \, z^2 = -1\}$; the cut is the projection on
the complex plane of the invariant variable $(z_1 - z_2)^2$ of pairs
of real points $(z_1,\ z_2)$ that have timelike or lightlike
separation. The analyticity of the two-point function in $\Delta$
therefore encodes the quantum mechanical requirement that field
operators at spacelike separations commute, while those at timelike
separations, in general, do not. The maximal analyticity property
has  been shown to be equivalent to a certain thermal KMS property
\cite{bros}. It is the closest analog of the analyticity property
implied by the positive energy condition for the case of Minkowskian
field theories. Other vacua (e.g. the so called alpha-vacua
\cite{Allen}) do not share these analyticity properties and have no
temperature.

For each (real or  complex) value of the mass parameter $m$ (or $\lambda$),
the relevant two-point function, solution of Eq (\ref{kg2p}), is
uniquely determined by the above analyticity properties, 
invariance, and the CCR;
it can be expressed in the complex domain in
terms of the hypergeometric function as follows

\begin{eqnarray} 
&& w_{\lambda}(z_1,\ z_2)  =
\Gamma(-\lambda) \; G_\lambda(\zeta)\ ,\ \ \ \zeta= z_1\cdot
z_2\,,\label{r.1}\\
&& \cr
&& G_\lambda(\zeta) = {\Gamma(\lambda+d-1)\over
(4\pi)^{d/2}\Gamma\left({d\over 2}\right)}\, 
F \left(-\lambda,\
\lambda+d-1;\ {d\over 2};\ {1-\zeta\over 2}\right )\ .
\label{r.1.1}\end{eqnarray} 
These two-point functions are closely
connected to the irreducible representations of the de Sitter group
and their spherical functions, studied in a number of mathematical
works, see e.g. \cite{GG,GGV,Fa,Mo,Tak,Di} and references therein.

\noindent $\bullet$ Theories in the above preferred family
corresponding to $m^2>0$ (and the limiting massless minimally
coupled case $m^2=0$) have been widely studied in the literature
with extremely important applications to inflation, and their
physical interpretation is still under active discussion
\cite{polyakov,polyakov2,bem1,bem2}.

\noindent $\bullet$ Theories corresponding to negative or, more
generally, complex squared masses give rise to quantum fields which
are local and de Sitter invariant but do not satisfy the
positive-definiteness condition;  one can reconstruct a linear
space of states  but the inner product
naturally associated with those two-point functions does not yield a
positive norm, and
therefore does not give rise to a Hilbert space; nor can a de Sitter
invariant positive subspace be found in the above linear space.
Therefore there is no acceptable quantum interpretation for the
field theories in this family.

There is however an exceptional family of masses
$m^2 = -n(n+d-1)$ corresponding to nonnegative integer
$ \lambda = n $ (in the $n=0$ case, we get the massless
minimally coupled field; see \cite{bcm} for a precursor of this paper), 
for which the two-point function is simply
infinite because of the pole of
$\Gamma(-\lambda)$ at $ \lambda = n $ in the RHS of Eq.~(\ref{r.1}); 
recall that the normalization of $w_\lambda$ in (\ref{r.1})
is determined by imposing the canonical commutation relations. 
On the other hand
\begin{equation}
F\left(-n,\ n+d-1;\ {d\over 2};\ {1-\zeta\over 2}\right ) =
\frac{\Gamma(n+1)\Gamma(d-1)}{\Gamma(n+d-1)}
C^{\frac{d-1}2}_n(\zeta)
\end{equation}
reduces to a Gegenbauer polynomial of degree $n$ (see \cite{HTF1});
it is a
holomorphic function without cuts and consequently with no quantum
feature.

We may try to give a meaning to this family of theories by
subtracting the divergent part. Since the pole of $\Gamma(-\lambda)$
is simple, with residue $(-1)^{n+1}/n!$, it is clear that the
following two-point function is well-defined:
\begin{equation}
\wh w_n(z_1,z_2)= \wh w_{n}(\zeta) = \lim_{\lambda\to n} \Gamma(-\lambda) \;
\left[G_\lambda(\zeta)-G_n(\zeta)\right] =\left . {(-1)^{n+1}\over
n!}{\partial \over \partial \lambda} G_\lambda(\zeta)\right
|_{\lambda = n}\ . 
\label{ren}\end{equation} 
Because $G_n(\zeta)$
has no discontinuity, the commutator $c_\lambda$ associated to
$w_\lambda$ (see (\ref{kgcomm})) tends to a well-defined limit as
$\lambda \rightarrow n$, without needing any subtraction, and this
limit is precisely the commutator associated to $\wh w_n$:
\begin{equation}
c_n(x,y) =  \lim_{\lambda\to n} c_\lambda(x,y) =  
\wh w_{n}(x,y)-\wh w_{n}(y,x)\ .
\label{rencom}\end{equation}
While the commutator solves the true Klein-Gordon equation, 
the two-point function $\wh w_n$ satisfies  a modified
tachyonic Klein-Gordon equation having an anomalous non-homogeneous
RHS as follows:
\begin{eqnarray}
&& \left[\Box  - n(n+d-1)\right] \widehat w_n(\zeta) =
{(-1)^{n+1}(2n+d-1)\Gamma \left ({d-1\over 2} \right ) \over
4\pi^{d+1\over 2}}
C^{\frac{d-1}2}_n(\zeta),
\label{eqmotion}
\end{eqnarray}
In the massless minimally coupled case the above equation becomes
simpler
\begin{eqnarray}
\Box \widehat w_0(\zeta) = -\frac{\Gamma\left({{d+1}\over
2}\right)}{2\pi^{\frac{d+1}{2}} };
\end{eqnarray}
the constant at the RHS is the inverse of the hypersurface of the unit 
sphere in dimension $d+1$. This is related to the fact that the Laplace
operator on the Euclidean de Sitter sphere has a zero mode and that
our subtraction scheme in this case amounts exactly to the removal
of that zero mode. In the procedure (\ref{ren}) leading to $\wh
w_n$, any meromorphic function of $\lambda$ with a simple pole at
$\lambda = n$ could have been used instead of $\Gamma(-\lambda)$:\
the result would have differed from $\wh w_n$ by the addition of a
multiple of $C^{\frac{d-1}2}_n(\zeta)$, without affecting
(\ref{eqmotion}).

Preserving a local and de Sitter invariant quantization
of this family of field theories
(including the massless minimally coupled scalar) is thus possible at the
expense of an anomalous non-homogeneous term in the quantum field
equations:
\begin{equation}
\left[\Box  - n(n+d-1)\right]\phi =  Q_n
\end{equation}
Of course we are really interested in (tachyonic) fields satisfying the correct
field equation with no non-homogeneous term.
We are therefore tempted to impose  the condition
\begin{equation}
Q_n^- \Psi = 0 \label{supp}
\end{equation}
on physical states, where $Q_n^-$ denotes the "annihilation" part 
of the operator $Q_n$.
An extraordinary property is that the above condition also selects a
{\emph positive} and {\emph de Sitter invariant} subspace of the space of
local states
and therefore opens the way for an acceptable quantum mechanical
interpretation of the  de Sitter tachyons (associated with the discrete
series of representations of the de Sitter group).
Full proofs will be published elsewhere. Here
we only give a hint of the proof of
this property in the one-particle subspace of the model. In this
case the condition is
\beq
\Psi \in \EE_n,\ \ \ \
\EE_n = \left\{ \Psi \in \CC_0^\infty(X_d)\ :
\int C^{\frac {d-1}2}_n(x_1\cdot x_2) \Psi(x_2) dx_2= 0\right\}
\endq
where $dx$ is the de Sitter invariant measure on
the hyperboloid $X_d$. $\EE_n$ is manifestly a de Sitter
invariant subspace of the one-particle local states.
The positivity in this subspace is a consequence of the following Fourier-type
(i.e. a "momentum space") representation of the two-point function 
$\wh w_n(z_1,z_2)$:
\beqa \wh w_{n}(z_1,\ z_2) &=&  \wh W_n(z_1,z_2)
-F^1_n(z_1,z_2) - F^2_n(z_1,z_2) +  b_n G_{n}(z_1,\ z_2). 
\label{s.7}\endqa

\begin{align} 
\wh W_n(z_1,z_2) &= a_n \int_\gamma\int_\gamma (z_1\cdot\xi)^{1-d-n}\,
(\xi\cdot \xi')^n\,\log(\xi\cdot \xi')\,
(z_2\cdot\xi')^{1-d-n}\,d\mu(\xi)d\mu(\xi') \label{physical} \\
F^1_n(z_1,z_2)&= a_n\int_\gamma\int_\gamma
\log(z_1\cdot\xi)(z_1\cdot\xi)^{1-d-n}\, (\xi\cdot \xi')^n\,
(z_2\cdot\xi')^{1-d-n}\,d\mu(\xi)d\mu(\xi')\\ 
F^2_n(z_1,z_2)&= a_n
\int_\gamma \int_\gamma
(z_1\cdot\xi)^{1-d-n}\, (\xi\cdot \xi')^n\,\log(z_2\cdot\xi')\,
(z_2\cdot\xi')^{1-d-n}\,d\mu(\xi)d\mu(\xi') 
\label{s.70}\end{align}
with
\beq
\ a_n = {(-1)^{n+1}\Gamma(n+d-1)^2\over 2^{n+2d}\pi^{3d-1\over
2}n!\Gamma\left(n+{d-1\over 2}\right)} \ \ \
b_n = {(-1)^{n+1}\over n!}{d\over d\lambda} \log
{\Gamma(\lambda+d-1)^2\over 2^{\lambda+2d}\pi^{3d-1\over
2}\Gamma\left(\lambda+{d-1\over 2}\right)} \bigg |_{\lambda = n }
\endq
Here  $\xi$ and $\xi'$ are vectors of the forward lightcone 
$C^+= \{\xi \in {{\bf R}^{d+1}}, \xi^2 = 0 , \, \xi^0>0\}$; 
the integrals are taken over the
section  $\gamma = C_+ \cap \{\xi\ :\ \xi^0 = 1\}$ i.e. over the
unit sphere in $\bR^{d}$; the measure $d\mu$ is the standard
invariant measure on the unit sphere.

Eq. (\ref{s.7}) is valid for any pair of complex de Sitter events 
$(z_1,z_2)$ such that $z_1$ belongs to the past tube $ \TT_-$ and 
$z_2$ to the future tube $ \TT_+$, defined as follows:
$\TT_\pm = \{z \in X_d^{(c)}, \Im z \in V^\pm\}$; 
this encodes precisely the  analyticity
property of the preferred vacua \cite{bros}. The two-point function 
$\wh w_n(x_1,x_2)$ is the boundary value on the reals from these tubes.

Note that the anomalous RHS in Eq. (\ref{eqmotion}) is due to the 
terms $F^1_n$ and $F^2_n$.
The proof that the  restriction of  the two-point function 
to ${\EE_n \times \EE_n}$ is de Sitter invariant
and positive-definite is based on the following  property:
if a test function $f$ belongs to ${\EE_n}$ the following integral vanishes
\begin{equation}
\int_{\gamma} d\mu(\xi) p(\xi) 
\int_{X_d}  (\xi \cdot (x\pm i0))^{1-n-d} \,  f(x) \,dx = 0
\end{equation}
for any polynomial $p$ of degree $\leq n$. 
This fact implies in particular that only the first term $\wh W_n(z_1,z_2)$
has a non-vanishing restriction to $\EE_n \times \EE_n $:
\beqa \wh w_{n}(z_1\cdot z_2)|_{\EE_n \times \EE_n}
 &=&  \wh W_n(z_1,z_2)|_{\EE_n \times \EE_n}. \label{oo}
\endqa
The proof of positive-definiteness starts from the fact that
$\xi\cdot \xi' = 1 - \vec \xi\cdot \vec \xi'$ for pair of points 
$\xi$ and $\xi'$  belonging
to the spherical section $\gamma$ of the cone.  By inserting  into Eq.
(\ref{physical}) the power series expansion 
\beq 
(-1)^{n+1}(1 - \vec
\xi\cdot \vec \xi')^n\, \log(1 - \vec \xi\cdot \vec \xi')/n! =
\sum_{m=0}^\infty u_{n,m}\, (\vec \xi\cdot \vec \xi')^m
\label{s.8.0}\endq 
in the variable $(\vec \xi\cdot \vec \xi')$, we get 
\begin{align} 
&\wh w_{n}(z_1\cdot z_2)|_{\EE_n \times \EE_n} =
{\Gamma(n+d-1)^2\over 2^{n+2d}\pi^{3d-1\over
2}\Gamma\left(n+{d-1\over 2}\right)} \times\cr
&\times \sum _{m= n +1 }^\infty u_{n,m}
\int_{\gamma\times \gamma} (z_1\cdot\xi)^{-n-d+1}\, (\vec \xi\cdot
\vec \xi')^m\,
(z_2\cdot\xi')^{-n-d+1}\,d\mu(\xi)\,d\mu(\xi')
\label{s.8}\end{align}
since terms of degree $\le n$ do not contribute in the physical
space. For $m> n$ the  coefficients $u_{n,m}$ are easily shown
to be positive real numbers; since the kernels $(\vec
\xi\cdot \vec \xi')^m$ in the above integrals are all
positive-definite, the positivity of the two-point function
restricted to the de Sitter invariant one-particle physical space
$\EE_n$ follows.

The Fock construction finally produces a local
and de Sitter covariant quantization of the tachyonic fields
corresponding to the squared masses $m^2 = -n(n+d-1)$. These
tachyons disappear in the flat limit. More precisely one can
consider the flat limit of the unrestricted two-point
function $\wh w_n(x_1\cdot x_2)$: the limit exists and coincides
with Schroer's two-point function, but the de Sitter invariant 
positive subspace gets
smaller and smaller (though infinite dimensional!) and  disappears
in the limit.

It is straightforward to obtain from our definition (\ref{ren})
an explicit expansion of $\wh w_n$ in terms of
$z= -(z_1-z_2)^2/4 = (1+\zeta)/2$.
In the case when $d\ge 2$ is an even integer, this expansion has
a simple structure:
\begin{equation}
\wh w_n(\zeta) = z^{1-{d\over 2}} A(z,\ n,\ d) -\log(z) B(z,\ n,\ d)
+ C(z,\ n,\ d),
\label{x.1}\end{equation}
where $A$, $B$, $C$ are polynomials in $z$:
\begin{align}
A(z,\ n,\ d) &= \sum_{m=0}^{{d\over 2}-2}
{z^{m}\Gamma \left ({d\over 2}-1-m \right)
\Gamma \left (n +{d\over 2}+m \right) \over
(4\pi)^{d\over 2}\Gamma(n +{d\over 2}-m)m!}\ ,\cr
B(z,\ n,\ d) &=
{\Gamma \left ({d-1\over 2}\right )\over
4 \pi^{d+1\over 2}}
C_n^{d-1\over 2}(1-2z)\ ,\cr
C(z,\ n,\ d) &=
\sum_{m=0}^n
{(-1)^mz^m \Gamma(n +d-1+m)\over
(4\pi)^{d\over 2} \Gamma(1+m) \Gamma\left ({d\over 2}+m \right)
\Gamma(n-m+1)}\times\cr
& \times \left [\psi(1+m) +\psi \left ({d\over 2}+m \right )
-2\psi(n+d-1+m) -\psi (1+n) \right ]\ .
\label{x.4}\end{align}
The most singular term in (\ref{x.1}) is
$(4\pi)^{-{d\over 2}} \Gamma\left( {d\over 2} -1 \right ) z^{1-{d\over 2}}$.
This reflects the normalization of $w_\lambda$, which is chosen
so as to satisfy the canonical commutation relations.
For example in dimension $d=4$,
\begin{equation}
\wh w_0(\zeta) = {1\over (4\pi)^2 z} - {2\over (4\pi)^2}\log(z)
- {4-2\gamma \over (4\pi)^2}\ .
\label{x.4.1}\end{equation}
This exhibits the local Hadamard behavior of this two-point function.
Only the first two terms in (\ref{x.1}) contribute to the commutator
$c_n(x_1,\ x_2)$:
\begin{eqnarray}
c_n(x_1,\ x_2) &=& \sum_{p=1}^{{d\over 2}-1}
{2^{2p+1}i\pi \Gamma(n+d-1-p)
\epsilon(x_1^0-x_2^0)\,\delta^{(p-1)}((x_1-x_2)^2)\over
(4\pi)^{d/2} \Gamma(n+1+p)\Gamma \left ({d\over 2}-p\right )}\cr
&& -i\epsilon(x_1^0-x_2^0) \theta((x_1-x_2)^2)
 {\Gamma \left ({d-1\over 2}\right )\over
2 \pi^{d-1\over 2}}
C_n^{d-1\over 2}(-x_1\cdot x_2)\
\label{x.5}\end{eqnarray}
In odd dimensions the expansion does not exhibit such features as polynomials
or logarithms. In all dimensions expansions in terms of the geodesic
distance $\sigma$ between $z_1$ and $z_2$ can be derived by substituting
$z = \sin^2(\sigma/2)$ and reexpanding. This would allow the computation
of what is referred to as the expectation value of the renormalized
stress-energy tensor, such as those of \cite{BuD,AF,BF,Tad}.
\vskip10pt

While  the above construction may be considered  satisfactory,
one can ask, following the usual way of understanding the massless minimally coupled field \cite{Allen,AF},
whether it would be possible to find a non-anomalous positive quantization
on the full test-function space $C^\infty_0(X_d) \times C^\infty_0(X_d)$.
Of course this quantization would necessarily break the de Sitter symmetry.
If it exists, the two-point function $W_n(x,y)$ solving this problem can be
decomposed into the de Sitter invariant part  $\wh w_n(x,y)$ plus a correction  $f_n(x,y)$:
\begin{eqnarray}
W_n(x,y) = \wh w_n(x,y)+ f_n(x,y).
\end{eqnarray}
1) The condition that the theory is canonical, i.e. that $W_n(x,y)$ has the right {\emph de Sitter invariant} commutator $c_n(x,y)$,
implies that  $f_n$ must be symmetric:
\beq
f_n(x,y) = f_n(y,x).
\endq
2) The absence of the anomalous term at the RHS of the Klein-Gordon equation
satisfied by $W_n$ is equivalent to the following non-homogeneous equations for $f_n(x,y)$:
\begin{align}
&\left[\Box_x  - n(n+d-1)\right] f_n(x,y) = 
\left[\Box_y  - n(n+d-1)\right] f_n(x,y)\cr
&= -
{(-1)^{n+1}(2n+d-1)\Gamma \left ({d-1\over 2} \right ) \over
4\pi^{d+1\over 2}}
C^{\frac{d-1}2}_n(\zeta);
\label{eqmotionbis}\end{align}
note the minus sign at the RHS. The second equation is actually 
implied by the symmetry of $f_n$ and of $C^{\frac{d-1}2}_n$.

\noindent 3)  $W_n(x,y)$ must be positive definite
to allow for the quantum mechanical interpretation also of 
the degrees of freedom
that break de Sitter invariance.

\noindent 4) If satisfied, the  condition
\begin{eqnarray}
 \left.f_n(x,y)\right|_{\EE_n \times \EE_n}=0
\end{eqnarray}
guarantees that the breaking of the de Sitter symmetry does not affect 
the  de Sitter invariant  physical subspace.

\vskip 10pt
In the massless minimally coupled case a partial solution to this problem 
satisfying condition 1) 2) and 3), but not 4),
is exhibited in \cite{Allen,AF}.
The explicit time dependence of the two-point function 
constructed there  has been given a physical interpretation
in the inflationary context and is regarded as a crucial 
feature of the model.

In the general case $n>0$, it is easy to find solutions that satisfy 
either the locality condition 1)
or the positive definiteness condition 3), but it does not seem possible 
to keep both properties
(in this connection see also \cite{Miao}).

In particular the standard KG equation holds by removing $F^1_n$ 
and $F^2_n$ from Eq. (\ref{s.7}).
To restore the positive definiteness
one needs to modify the coefficients of the first $n$ terms 
in the series (\ref{s.8.0}).
The resulting two-point function satisfies the Klein-Gordon equation 
and still coincides
with $\wh w_n$ on ${\EE_n \times \EE_n}$. However it is not canonical
and its usefulness outside of the physical subspace appears dubious.

The existence of the above mentioned partial  solution to the  
massless minimal coupled
model is due to the fact that this theory is also a limiting case 
of the positive
squared mass theories.  However, from the present point of view,
the correct massless minimally coupled field is better described by
\begin{equation}
W_0(x_1,x_2)|_{\EE_0 \times \EE_0}= \wh w_0(x_1\cdot x_2)|_{\EE_0 \times \EE_0}
\end{equation}
which is a local de Sitter invariant quantization of
that field on the space of test functions having zero mean value and 
is a special instance of our general construction.

We also note that the fields which we have constructed are related
to those constructed on the sphere $S_d$ (i.e. the ``Euclidian''
version of $X_d$) by A.~Folacci in \cite{Fo}. However this formalism
makes it difficult to study the physical space and the positivity
of the metric there.

\end{document}